\documentclass[a4paper,fleqn,usenatbib]{mnras}
\usepackage{graphicx}
\usepackage{amsmath}
\usepackage{amssymb}
\usepackage{mathrsfs}

\title{On neutralization of charged black holes}

\author[Gong, Cao, Gao \& Zhang]{
Yi Gong,$^{1}$
Zhoujian Cao,$^{2}$\thanks{E-mail:zjcao@amt.ac.cn}
He Gao,$^{2}$
and Bing Zhang,$^{3}$
\\
$^{1}$School of Physical Science and Technology, Southwest Jiaotong University, Chengdu 610031, China\\
$^{2}$Department of Astronomy, Beijing Normal University, Beijing 100875, China\\
$^{3}$Department of Physics and Astronomy, University of Nevada Las Vegas, NV 89154, USA
}

\date{Accepted XXX. Received YYY; in original form ZZZ}

\pubyear{2019}

\begin{document}
\label{firstpage}
\pagerange{\pageref{firstpage}--\pageref{lastpage}}
\maketitle

\begin{abstract}
For non-spinning, charged (Reissner-Nordstr\"om) black holes,
the particles with an opposite sign of charge with respect to that of the black hole will be pulled into the black hole by the extra electromagnetic force. Such a hole will be quickly neutralized so that there should not exist significantly charged, non-spinning black holes in the universe. The case of spinning, charged (Kerr-Newmann) black holes is more complicated. For a given initial position and initial velocity of the particle, an oppositely charged particle does not always more easily fall into the black hole than a neutral particle. The possible existence of a magnetosphere further complicate the picture. One therefore cannot straightforwardly conclude that a charged spinning black hole will be neutralized. In this paper, we make the first step to investigate the neutralization of Kerr-Newmann black holes without introducing a magnetosphere. We track the particle trajectories under the influence of the curved spacetime and the electromagnetic field carried by the spinning, charged black hole. A statistical method is used to investigate the neutralization problem. We find a universal dependence of the falling probability into the black hole on the charge of the test particle, with the oppositely charged particles having a higher probability of falling. We therefore conclude that charged, spinning black holes without a magnetosphere should be quickly neutralized, consistent with people's intuition. The neutralization problem of Kerr-Newmann black holes with a co-rotating force-free magnetosphere is subject to further studies.
\end{abstract}


\section{Introduction}
Right after the announcement of the detection of the first gravitational wave event \citep{Abbott:2016blz} and its putative $\gamma$-ray counterpart \citep{connaughton16},  \cite{Bing2016MERGERS} proposed a possible mechanism of producing a brief electromagnetical counterpart signal of binary black hole merger gravitational wave events. The key requirement of Zhang's mechanism is that at least one of the binary black hole members admits electric charge. It has been widely believed that the amount of charge carried by astrophysical black holes is negligible.  \cite{Bing2016MERGERS} argued that a spinning charged black hole may carry a force-free magnetosphere which may sustain charge for an extended period of time. Later, several investigations of electromagnetic counterparts of binary black hole mergers or fast radio bursts also invoked electric charges in black holes \citep{liebling16,liu16,MN459L41,fraschetti18,levin18,deng18}, and it has been shown that Kerr-Newmann black holes can be formed within the astrophysical context through directly collapsing spinning magnetized neutron stars \citep{nathanail17}. 
On the observational side, constraints on the amount of charge carried by the Sagittarius A* black hole has been carried out, and the data cannot rule out the existence of some charge from the black hole \citep{Zajacek:2018vsj,Zajacek:2019kla}. It is therefore highly interesting to investigate the neutralization problem of charged black holes.

The black hole neutralization problem has been investigated by several authors in the past. \cite{EardleyPress75} presented the following estimation: For a representative particle with mass $m_p$ (mass of a proton) and charge $e$ (charge of a electron), the acceleration from the electric force of the charged black hole is proportional to $\frac{e}{m_p}Q$, where $Q$ is the charge of the black hole. The acceleration from the gravity of the charged black hole, on the other hand, is proportional to $M$, where $M$ is the mass of the black hole. So, if $\frac{e}{m_p}Q>M$, or equivalently $\frac{Q}{M}>\frac{m_p}{e}\sim10^{-18}$, the electric force would dominate the gravitational force to let the black hole attract more particles with the opposite sign of charge. Based on this simple argument, \cite{EardleyPress75} suspected that no charged black hole with $\frac{Q}{M}$ greater than $10^{-18}$ exists in nature. \cite{PhysRevD.18.3598} discovered that a black hole may result in eddy currents. These currents would possibly affect the motion of charged particle.  \cite{RUFFINI20101} discussed the vacuum polarization process near a charged black hole which would produce an electromagnetic environment that may affect the discharge problem of the black hole. \cite{PhysRevD.96.063015} constructed a simplified model for accretion of charged particles by a Kerr-Newman black hole. They found that a small amount of charge of the black hole may in general have a non-negligible effect on the motion of the plasma, as long as the electromagnetic field of the plasma is still negligible.

Intuitively, most people think charged black holes are unrealistic, since they tend to attract opposite charges from the ambient medium to neutralize themselves. Such an intuitive thinking can be easily proved for charged black holes with a spherical symmetry, i.e. without spin. For such a Reissner-Nordstrom (RN) black hole, both the gravitational force and the electromagnetic force are in the radial direction, so the particles with an opposite charge with respect to the black hole will be more easily pulled into the black hole than neutral particles and the particles with the same sign of charge with respect to the black hole. We may therefore conclude that a Reissner-Nordstrom black hole can be essentially neutralized within a short period of time.

If the charged black hole is spinning, the situation becomes much more complicated. Even if one does not consider the possibility of a force-free magnetosphere surrounding the spinning hole, the trajectory of a particle in the vicinity of a Kerr-Newman (KN) black hole cannot be described intuitively. The black hole's spin produces a magnetic component of the gravitational force on the test particle. The direction of this force component is not along the radial direction, but is related to the velocity of the test particle. Further complication for a spinning black hole is the existence of the ergosphere. Within the ergosphere, orbits with negative energies exist. This is a pure general relativistic effect without a Newtonian counterpart.

With the nontrivial spin effect, the intuitive thinking and argument on the neutralization of RN black holes is not valid for KN black holes. This motivates us to quantitatively investigate the neutralization problem of KN black holes. A KN black hole likely carries a force free magnetosphere, which makes the neutralization problem very complicated. As the first step, we neglect all the relevant plasma processes without introducing a magnetosphere, but only track the trajectories of individual particles under the influence of the curved spacetime and the electromagnetic field carried by the spinning, charged black hole. Our purpose is to investigate whether oppositely charged particles with respect to the hole has a higher probability falling into the hole. Since for a given initial position and initial velocity of the particle, an oppositely charged particle does not always more easily fall into the black hole than a neutral particle, we use a statistical method to do the investigation, which is different from the method previously used in the literature \citep[e.g.][]{PhysRevD.87.124030}.

The structure of the paper is as follows. In Sec.~\ref{secII}, we present the dynamical equations to delineate the motion of charged test particles around a Kerr-Newman black hole. An analysis is performed to introduce a convenient method to study the neutralization problem. In Sec.~\ref{secIII}, we introduce a statistical method to treat the neutralization problem. The orbits of test particles are determined by their initial positions and velocities, but can also be described by the conserved quantities including energy, angular momentum, and the Carter constant. We apply a Monte Carlo method by randomly producing test particles with different initial conditions and sign of charge and check their distributions with respect to these conserved quantities. With the information, we can investigate the probabilities of the three types of particles (neutral and those with the same or opposite charge with respect to the black hole) falling into the black hole in the Sec.~\ref{secIV}. The results are summarized in Sec.~\ref{secV}.

We adopt the units with $c=G=4\pi\epsilon_0=\frac{4\pi}{\mu_0}=1$ throughout the paper, where $\epsilon_0$ and $\mu_0$ are the electric permittivity and magnetic permeability of the vacuum, respectively \citep{liang00}.

\section{Dynamical equations of charged particles in Kerr-Newman spacetime}\label{secII}
The dynamical equations of charged particles in the KN spacetime can be written as \citep{liang00}
\begin{align}
\label{dynamics}
mU^\nu \nabla_\nu U^\mu=qU^\nu F^\mu{}_\nu,
\end{align}
where $q$ and $m$ are the charge and mass of the particle, $U^\mu$ is its four velocity in the spacetime, $\nabla$ and $F_{\mu\nu}$ are the covariant derivative operator and the electromagnetic field tensor of the KN spacetime, respectively. There are four constants of motion for this dynamics \citep{MTW}. They are energy $\mathscr{E}$ measured at infinity, extended angular momentum $\mathscr{L}_z$, length of the four velocity $\kappa$ and the Carter constant $\mathscr{K}$. They can be expressed as
\begin{align}
\mathscr{E}&\equiv -mg_{t\mu}U^\mu-qA_t,\label{energy}\\
\mathscr{L}_z&\equiv mg_{\phi\mu}U^\mu+qA_\phi,\label{Amomentum}\\
\kappa&\equiv g_{\mu\nu}U^\mu U^\nu=-1,\label{kappa2}\\
\mathscr{K}&\equiv(mg_{\theta\mu}U^\mu)^2+\cos^2\theta[a^2(m^2-E^2)+(\mathscr{L}_z/\sin\theta)^2],\label{carter}
\end{align}
where $g_{\mu\nu}$ is the metric in the Boyer-Lindquist coordinate $(t,r,\theta,\phi)$, and $a$ is the spin parameter of the KN black hole. Since $m$ and $q$ are constants, only the ratio of charge and mass $q/m$ (specific charge $\eta\equiv q/m$) affects the dynamics. So the system we consider includes four physical parameters $(M, a, Q, \eta)$ where $M$ and $Q$ are the mass and the charge of the KN black hole, respectively. Given a set of physical parameters $(M, a, Q, \eta)$, the orbit of a charged particle is determined by the initial conditions $(t_0, r_0, \theta_0, \phi_0, U^t_0, U^r_0, U^\theta_0, U^\phi_0)$, where $U^\mu\equiv\frac{dx^\mu}{d\tau}$ with $\tau$ being the proper time along the world line of the particle's orbit. Since the KN spacetime is stationary and axisymmetric, we can always set the Boyer-Lindquist coordinate to make $t_0=\phi_0=0$.

Within the Boyer-Lindquist coordinate system, the above four constants of motion can be expressed as \citep{PhysRevD.87.124030}
\begin{align}
&E\equiv\mathscr{E}/m=-g_{tt}\frac{dt}{d\tau}-g_{t\phi}\frac{d\phi}{d\tau}-\eta A_t,\\
&L_z\equiv\mathscr{L}_z/m=g_{t\phi}\frac{dt}{d\tau}+g_{\phi\phi}\frac{d\phi}{d\tau}+\eta A_\phi,\\
&\kappa=-1,\\
&K\equiv\mathscr{K}/m^2=(g_{\theta\theta}\frac{d\theta}{d\tau})^2\nonumber\\
&+\cos^2\theta\{a^2[1-(\frac{\mathscr{E}}{m})^2]+(\frac{\mathscr{L}_z}{m})^2\frac{1}{\sin^2\theta}\}\label{Keq}
\end{align}
Based on these constants of motion $(\mathscr{E},\mathscr{L}_z,\kappa,\mathscr{K})$, the equations of motion can be written as (Eqs.~(33.32) of \citep{MTW})
\begin{align}
&(m\Sigma\frac{dr}{d\tau})^2=\mathscr{R},\\
&(m\Sigma\frac{d\theta}{d\tau})^2=\Xi,\\
&m\Sigma\frac{d\phi}{d\tau}=-(a\mathscr{E}-\frac{\mathscr{L}_z}{\sin^2\theta})+\frac{a}{\Delta}\mathscr{P},\\
&m\Sigma\frac{dt}{d\tau}=-a(a\mathscr{E}\sin^2\theta-\mathscr{L}_z)+\frac{(r^2+a^2)}{\Delta}\mathscr{P},
\end{align}
where $\Delta\equiv r^2-2Mr+a^2+Q^2,\Sigma\equiv r^2+a^2\cos^2\theta$ and
\begin{align}
&\mathscr{R}=\mathscr{P}^2-\Delta[m^2r^2+(\mathscr{L}_z-a\mathscr{E})^2+\mathscr{K}],\\
&\Xi=\mathscr{K}-\cos^2\theta[a^2(m^2-\mathscr{E}^2)+\frac{\mathscr{L}^2_z}{\sin^2\theta}],\\
&\mathscr{P}=(r^2+a^2)\mathscr{E}-a\mathscr{L}_z-qQr,\\
&A_t=-\frac{Qr}{\Sigma},\\
&A_\phi=\frac{Qr}{\Sigma}a\sin^2\theta.\label{EMforcephi}
\end{align}

We can reduce the above equations with $m$ to get
\begin{align}
&(\Sigma\frac{dr}{d\tau})^2=R,\\
&(\Sigma\frac{d\theta}{d\tau})^2=\Theta,\\
&\Sigma\frac{d\phi}{d\tau}=-(aE-\frac{L_z}{\sin^2\theta})+\frac{a}{\Delta}P,\\
&\Sigma\frac{dt}{d\tau}=-a(aE\sin^2\theta-L_z)+\frac{(r^2+a^2)}{\Delta}P,
\end{align}
where $\Delta\equiv r^2-2Mr+a^2+Q^2,\Sigma\equiv r^2+a^2\cos^2\theta$ and
\begin{align}
&R=P^2-\Delta[r^2+(L_z-aE)^2+K],\label{Rfunction}\\
&\Theta=K-\cos^2\theta[a^2(1-E^2)+\frac{L^2_z}{\sin^2\theta}],\\
&P=(r^2+a^2)E-aL_z-\eta Qr.
\end{align}

Apparently $R$ depends on $r$, $E$, $L_z$, $K$, $M$, $a$, $Q$ and $\eta$. For a given black hole and a charged particle, $M$, $a$, $Q$ and $\eta$ are fixed. If the initial condition of the charged particle is specified,
$E$, $L_z$ and $K$ are also determined. Then we get the function $R(r)$. Only positions with $R(r)\geqslant0$ the test particle can access. The condition $R(r)\geqslant0$ divides the space outside of the black hole horizon into several disconnected regions. If the initial position and the black hole horizon are located in the same region, the particle will fall into the black hole. If the initial position, the black hole horizon and $r=\infty$ are all located in the same region, particles with $U^r_0<0$ will fall into the black hole. Otherwise the test particle will stay outside of the black hole. As a result, for a given black hole and a charged particle with the initial information $(t_0, r_0, \theta_0, \phi_0, U^t_0, U^r_0, U^\theta_0, U^\phi_0)$, one can easily judge whether the particle will fall into the black hole based on the analysis of the behavior of $R$. This trick has been used before in \citep{PhysRevD.87.124030,YangWang13}.

\begin{figure}
\begin{tabular}{c}
\includegraphics[width=0.5\textwidth]{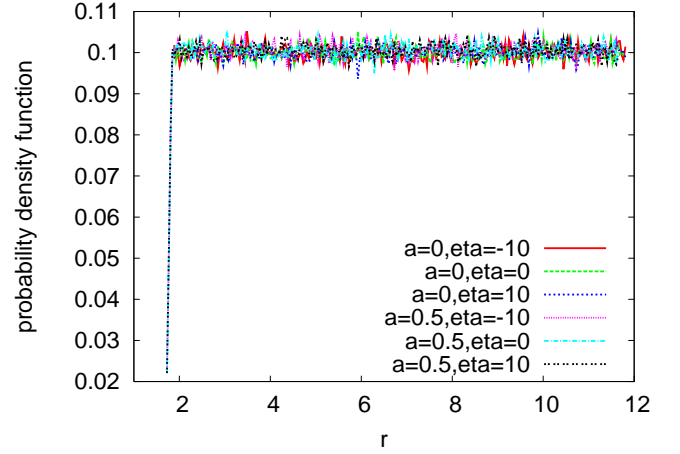}
\end{tabular}
\caption{The probability density function of the initial state distribution with respect to $r$. The parameters $Q=0.5$, $r_{max}=10$ and $v_{max}=1$ are adopted.}\label{fig1}
\end{figure}
\begin{figure*}
\begin{tabular}{cc}
\includegraphics[width=0.5\textwidth]{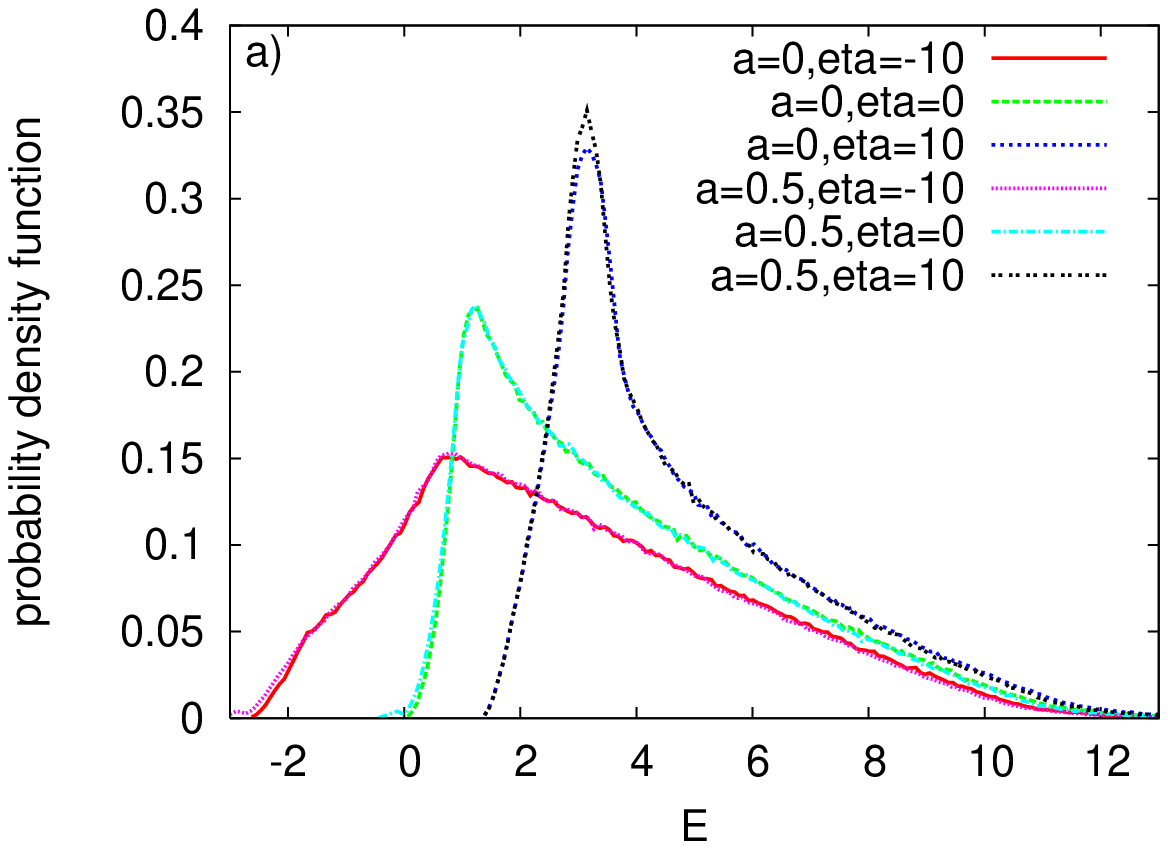}&
\includegraphics[width=0.5\textwidth]{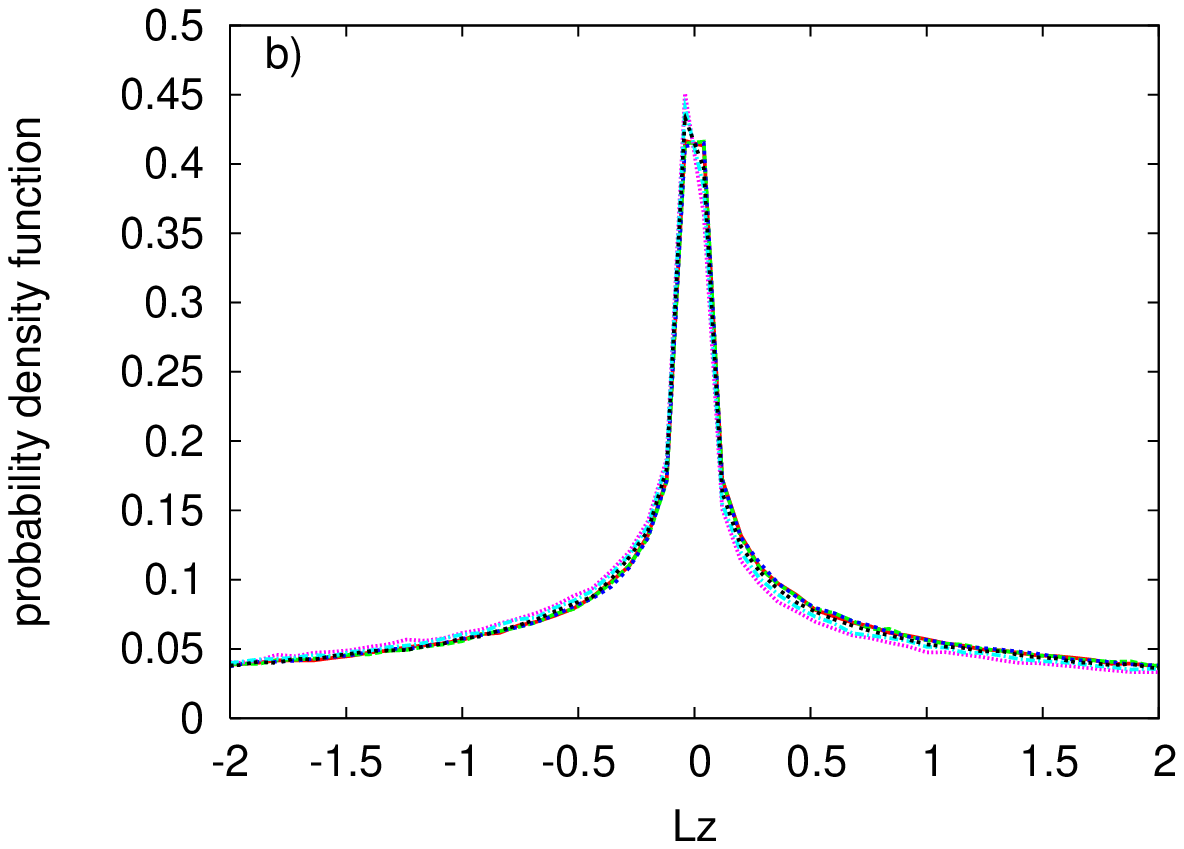}\\
\includegraphics[width=0.5\textwidth]{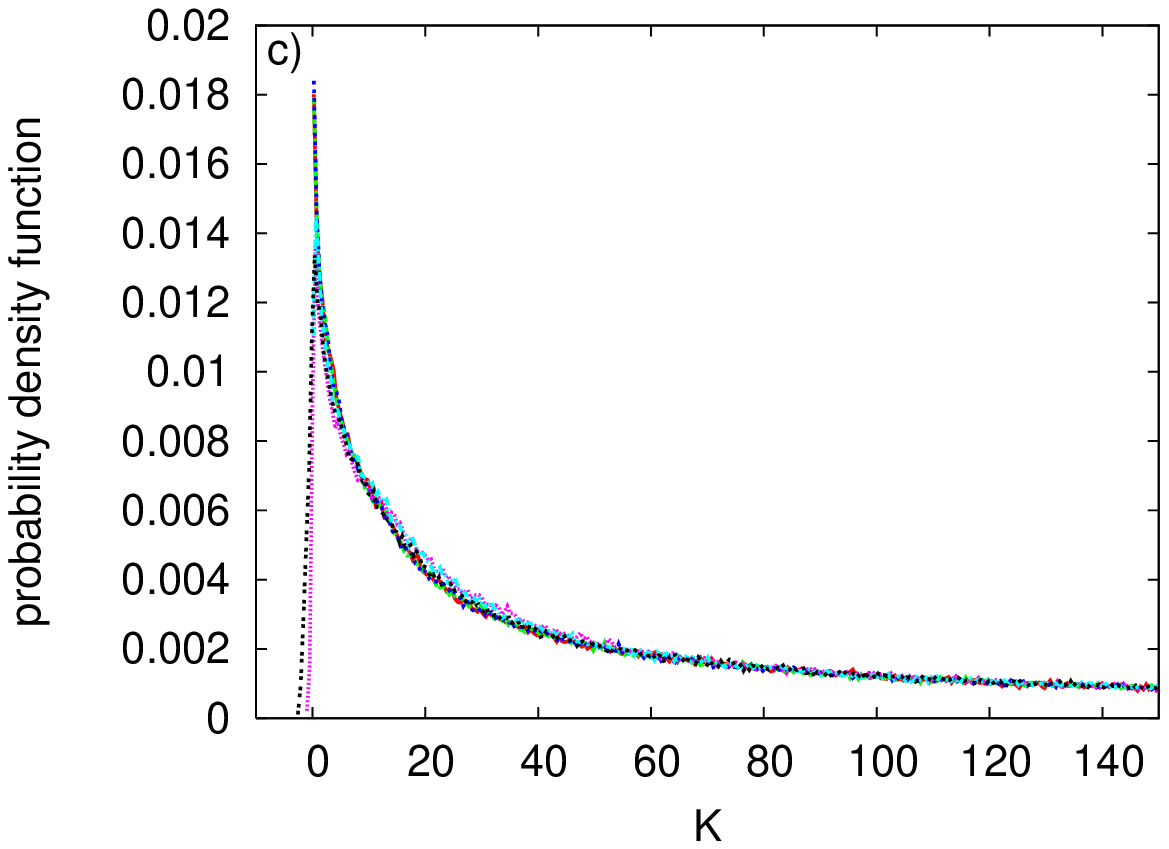}&
\includegraphics[width=0.5\textwidth]{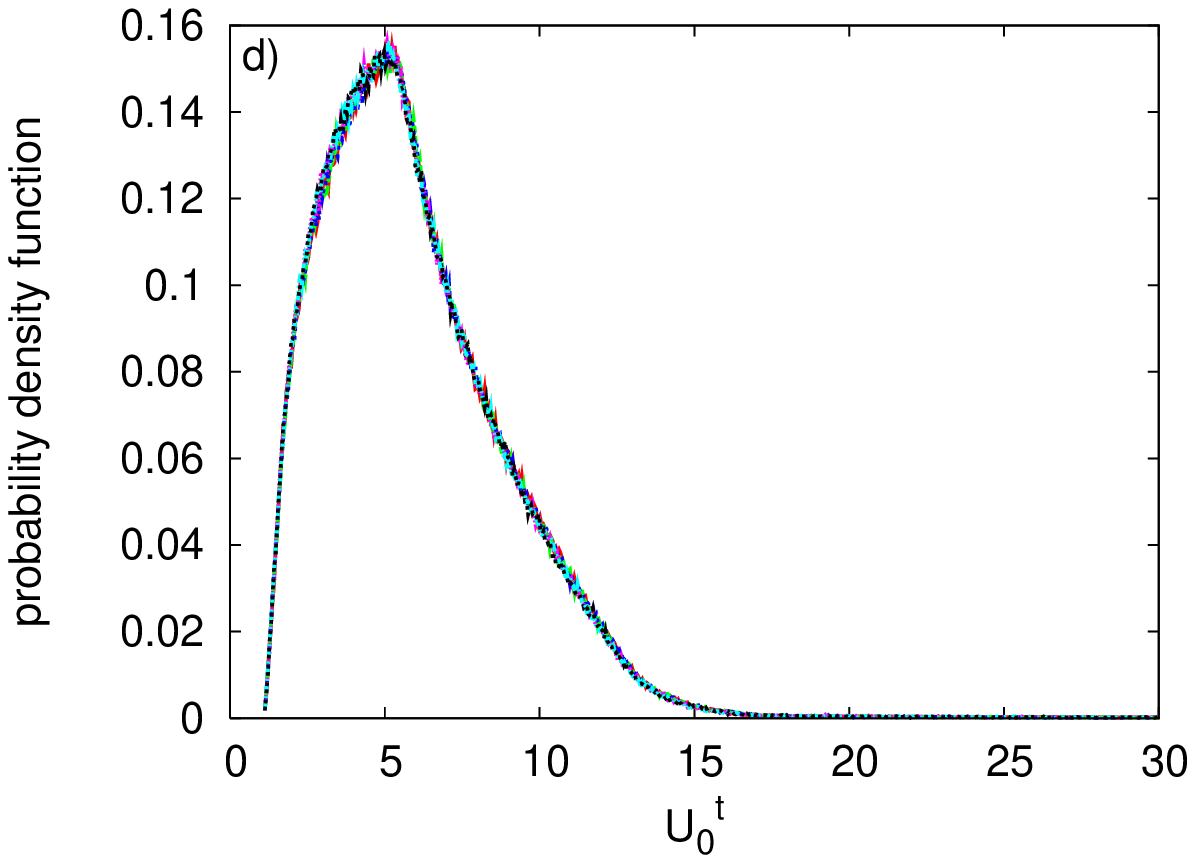}
\end{tabular}
\caption{The probability density function of the initial state distributions with respect to $E$, $L_z$, $K$ and $U^t_0$. The parameters $Q=0.5$, $r_{max}=10$ and $v_{max}=1$ are adopted.}\label{fig2}
\end{figure*}

\section{Statistical properties of the initial conditions}\label{secIII}
Since the trajectories of the particles depend on the initial conditions, and since for a given initial condition, an oppositely charged particle may not always more easily fall into the black hole, we perform a Monte Carlo simulation to statistically investigate the probability of each type of particle (neutral, with the same or opposite sign of charge) falling into the hole. Given a same set of black hole parameters $(M, a, Q)$, we investigate three types of particles: $\eta = -10, 0, 10$, which represent charge with opposite sign with respect to the hole, neutral, and charge with the same sign with respect to the hole. The absolute value of $|\eta|=10$ is chosen arbitrarily for the convenience of our investigation. The unit we use corresponds to the International System of units (SI) through $\frac{q}{m}=\frac{q_{SI}}{m_{SI}\sqrt{4\pi G\epsilon_0}}$. Adopting $\epsilon_{0}=8.85\times10^{-12}$Fm${}^{-1}$, $G=6.67\times10^{-11}$m${}^3$kg${}^{-1}$s${}^{-2}$ in SI, an electron corresponds to $\eta=2.04\times10^{21}$ and a proton corresponds to $\eta=1.11\times10^{18}$. Macroscopic charged clumps should admit smaller $\eta$ values than electron and proton. In the current work, we adopt $|\eta | =10$, but different absolute values of $\eta$ does not affect the statistical results presented below.

Due the time symmetry and the rotation symmetry of the spacetime, $t_0$ and $\phi_0$ do not play an essential role in our problem, so we can neglect these two parameters. For $\theta_0$, we assume that it uniformly takes random values from $(0,\pi)$. For $r_0$, we assume that it uniformly takes random values from $(r_{0+},r_{max})$ where $r_{0+}$ is the radius of the outer horizon of the black hole, and $r_{max}$ is an arbitrary maximum radius in our consideration, the specific value of which does not affect our results. Regarding the initial velocity, $U^t_0$ and $U^r_0$ have the dimension $1$, while $U^\theta_0$ and $U^\phi_0$ have the dimension $1/M$. From now on in the current paper, we take $M=1$ without losing generality. So we can assume $U^r_0$, $U^\theta_0$ and $U^\phi_0$ all uniformly take random values from the range $(-v_{max},v_{max})$, where $v_{max}$ is an arbitrary maximum value of velocity, the specific value of which does not affect our results. The remaining $U^t_0$ is determined by the condition $\kappa=-1$. One important issue is that the non-trivial spacetime structure may make some values of $(U^r_0, U^\theta_0, U^\phi_0)$ unphysical, so that $\kappa=-1$ cannot be satisfied in those cases. These trial values are excluded from our simulation.

Due the nonlinear relation between the initial conditions and the conserved quantities, $E$, $L_z$ and $K$ are not uniformly distributed. In the following, we investigate the statistical behavior of $E$, $L_z$ and $K$. $U^t_0$ is related to the energy detected by the comoving observer, which is strongly correlated to $E$.

More concretely, in our setting the initial states are uniformly distributed with respect to $r$, $\theta$, $U^r_0$, $U^\theta_0$ and $U^\phi_0$. One subtlety is that some combinations of $r$, $\theta$, $U^r_0$, $U^\theta_0$ and $U^\phi_0$ may be ruled out by the requirement of $\kappa=-1$. So the resulting initial states may differ from the uniform distribution with respect to $r$. In Fig.~\ref{fig1}, we can see that such deviation may occur for the positions inside of the ergosphere. Note that the outer horizon is at $r_{0+}=M+\sqrt{M^2-Q^2-a^2}$, which is also the inner boundary of ergosphere. The outer boundary of ergosphere, on the other hand, is at $M+\sqrt{M^2-Q^2-a^2\cos^2\theta}$, which is $\theta$-dependent. For the $a=0$ case, the uniform distribution is well preserved, and the $r$ range that is valid is $r>r_{0+}\approx1.866$. For the $a>0$ case (e.g. $a=0.5$ in Fig.~\ref{fig1}), on the other hand, even though the uniform distribution is also satisfied outside the ergosphere $r \gtrsim 1.866$, a deviation from the uniform distribution shows up for the positions inside the ergosphere (the rising part of the probability distribution). The smallest radius is at $r_{0+} (\theta=0) \approx 1.707$ for $a=0.5$. Since the ergosphere region increases with increasing $\theta$, the ergosphere effect becomes progressively important from 1.707 to 1.866, until reaching the uniform distribution above 1.866. The deviation below $r \approx 1.866$ is independent of particle's charge.

Hereafter we have used the probability density function to quantitatively describe the behavior. The ``probability'' is defined as the ratio $F\equiv\frac{N_{\text{pr}}}{N_{\text{tot}}}$ of our Monte Carlo simulations, where $N_{\text{tot}}$ is the number of all the simulated samples and $N_{\text{pr}}$ is the number of samples with the parameter(s) falling in the given range. The probability density function is defined as $\frac{F}{\Delta}$ where $\Delta$ is the range of the given parameter. For example, in Fig.~\ref{fig1} where the parameter is $r$, we have taken the parameter range as $\Delta=0.04$. We have taken a small enough range $\Delta$ to allow the numerical error not to affect the density function.

The energy $E$ includes two parts: one is the mechanical energy $-g_{t\mu}U^\mu$ and the other is the electromagnetic energy $-\eta A_t$, which is related to the particle's charge. Due to the future-directed time-like property of $U^\mu$, the mechanical energy part is always positive when the particle is outside the ergosphere. As a result, we can see that even though a neutral particle in Schwarzschild spacetime ($a=0$ and $\eta=0$ in the Fig.~\ref{fig2}a) allows positive energy only, negative energy states can also appear for neutral particles in Kerr spacetime ($a=0.5$ and $\eta=0$ in the Fig.~\ref{fig2}a). Since the volume of the ergosphere is relatively small, the correction introduced by negative energy states with respect to Schwarszchild spacetime is small. On the other hand, the sign of the electromagnetic energy is determined by the sign of the particle's charge, i.e. positive for $\eta>0$ and negative for $\eta < 0$. As a result, a positive $\eta$ shifts the energy distribution towards the positive side and vice versa.

Like energy $E$, the angular momentum also includes the mechanical part and the electromagnetic part. The mechanical part is affected by the rotation of the black hole. With respect to the Boyer-Lindquist coordinate used in our calculation, the zero angular momentum orbit allows a positive $U^\phi$. So our uniform distribution with respect to $U^\phi$ results in a $L_z$ that is most probably negative, as shown in Fig.~\ref{fig2}b. In contrast, the electromagnetic part allows a spherical symmetric behavior. This explains why when $\eta$ increases, the $L_z$ distribution behaves more and more like the $a=0$ case. When $a=0$, different $\eta$ cases make little difference in the $L_z$ distribution.

From Eq.~(\ref{Keq}), one can see that only when $E<1$ and when $E$ term dominates will one have $K<0$. Otherwise $K>0$ is satisfied. So Fig.~\ref{fig2}c indicates that most initial states allow positive $K$ values. In addition, we can see that the black hole spin and the particle's charge introduce little change in the $K$ distribution.

Finally, $U^t_0$ is constrained by the requirement $\kappa=-1$. $U^t_0$ is similar to the mechanical energy but less affected by the ergosphere. As a result, one can see that the distribution form is similar to the $E$ distribution with $a=\eta=0$ case. And the distribution is marginally affected by $a$ and $\eta$ as shown in Fig.~\ref{fig2}d.

In Figs.~\ref{fig1} and \ref{fig2}, we have used $Q=0.5$, $r_{max}=10$ and $v_{max}=1$. Adopting other values would only change the scales of the plots without changing the overall behavior.

\section{Probabilities of charged particles falling into the black hole}\label{secIV}
\begin{figure*}
\begin{tabular}{cc}
\includegraphics[width=0.5\textwidth]{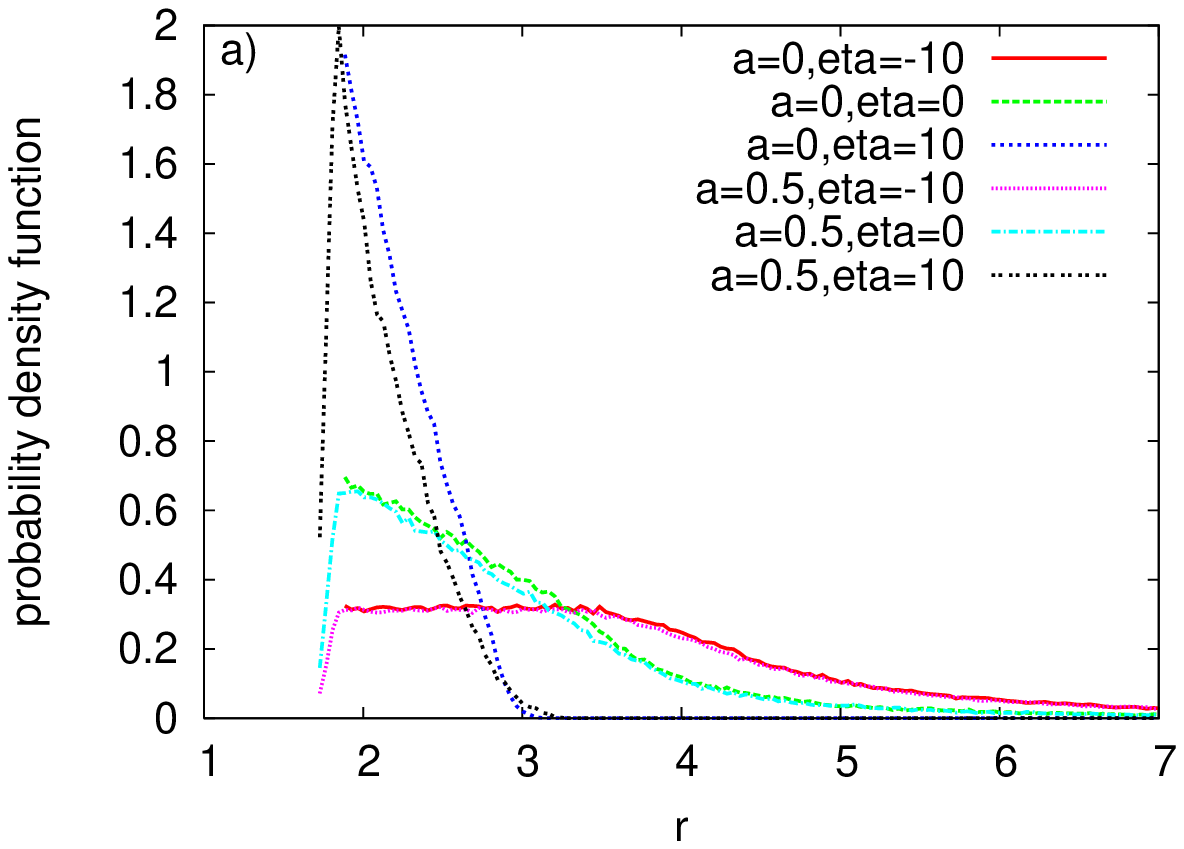}&
\includegraphics[width=0.5\textwidth]{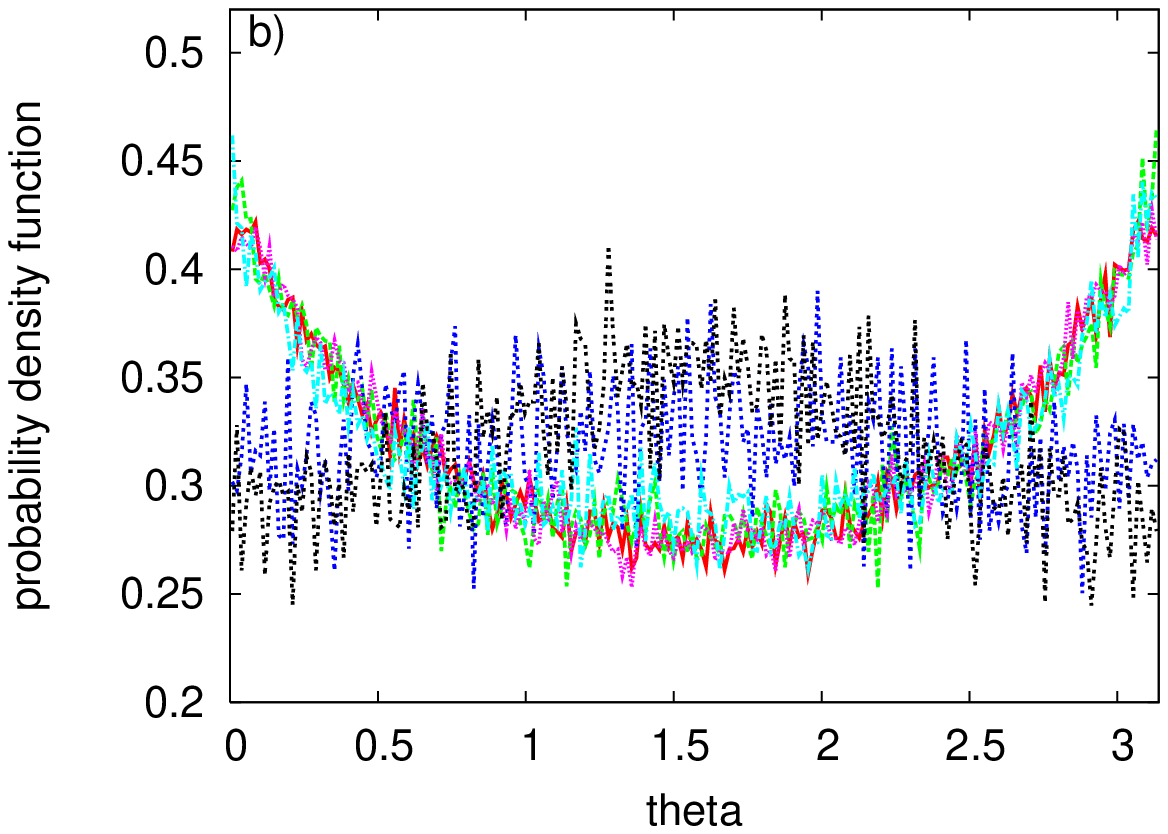}
\end{tabular}
\caption{The probability density function for charged particles falling into black hole with respect to the initial position coordinates $r$ and $\theta$. The parameters $Q=0.5$, $r_{max}=10$ and $v_{max}=1$ are adopted.}\label{fig3}
\end{figure*}

In this section, we investigate the probabilities for the particles with different charges falling into the black hole. Based on the initial condition distributions discussed in Sect.\ref{secIII}, we investigate how the falling probability depends on the particle charge and the parameters of the black hole.

First, we investigate the cases with fixed $a$, $Q$ and $\eta$ values. In reference of Fig.~\ref{fig2}, we consider the probability for charged particles falling into black hole with respect to the initial position, initial velocity and the conserved quantities. Here the probability means the fraction for particles falling into the black hole with a specific initial position range, for example, relative to the case with any initial positions.

\begin{figure}
\begin{tabular}{c}
\includegraphics[width=0.5\textwidth]{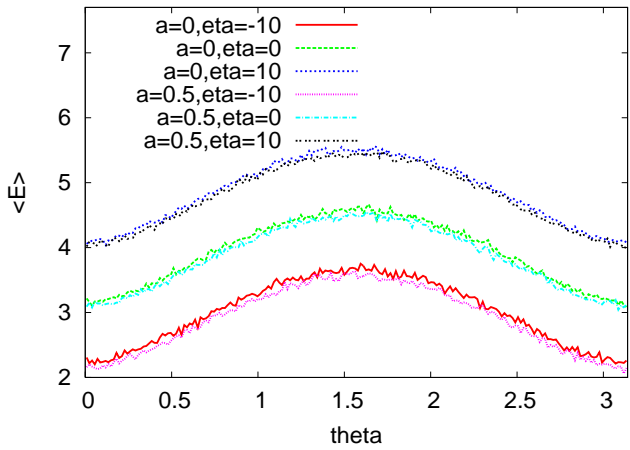}
\end{tabular}
\caption{The averaged energy for the initial state particles as a function of $\theta$. The parameters $Q=0.5$, $r_{max}=10$ and $v_{max}=1$ are adopted.}\label{fig4}
\end{figure}

\begin{figure*}
\begin{tabular}{ccc}
\includegraphics[width=0.333\textwidth]{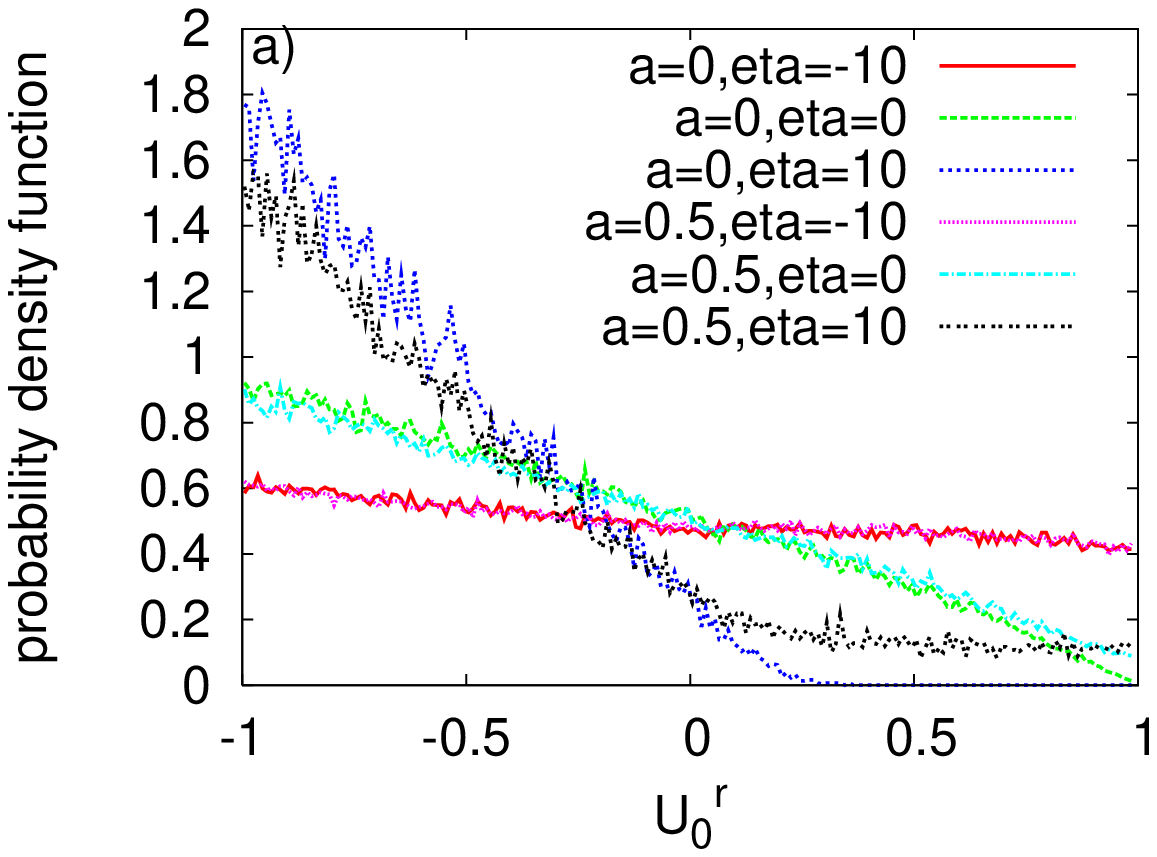}&
\includegraphics[width=0.333\textwidth]{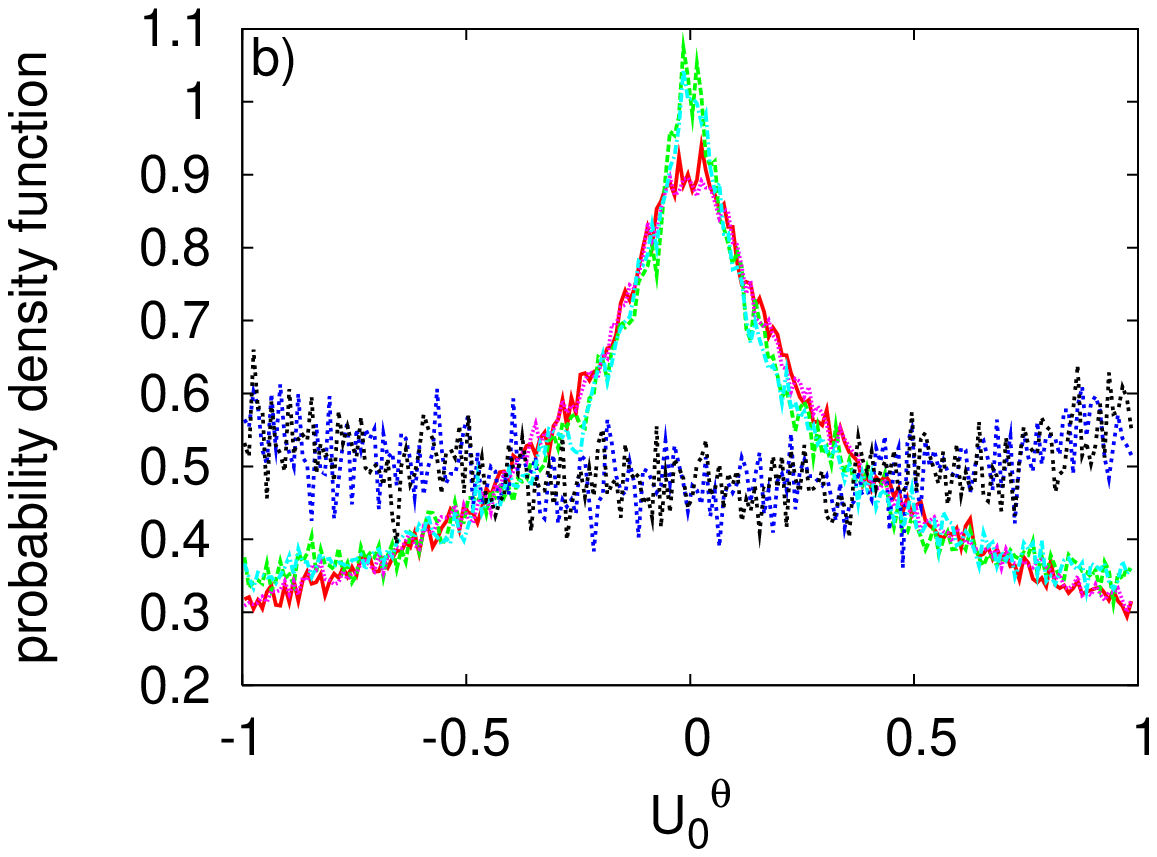}&
\includegraphics[width=0.333\textwidth]{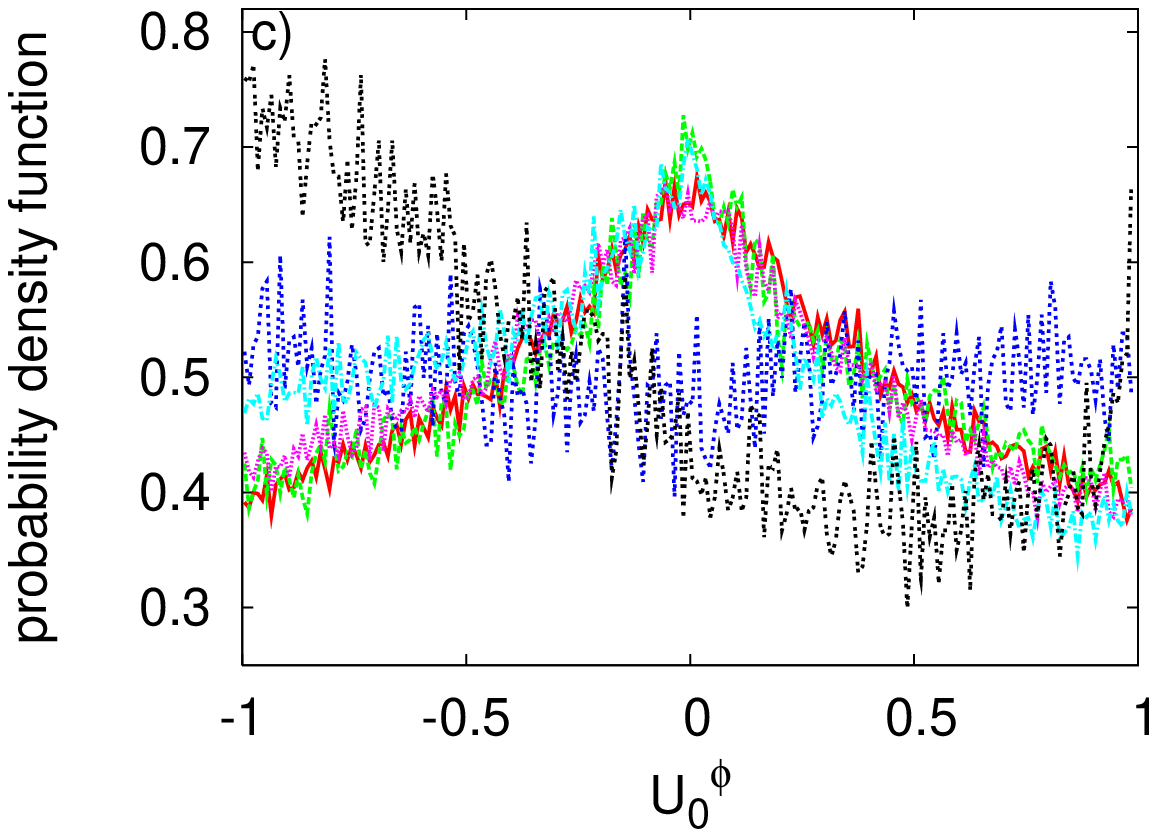}
\end{tabular}
\caption{The probability density function for particles falling into black hole with respect to the three components of the initial velocities. The parameters $Q=0.5$, $r_{max}=10$ and $v_{max}=1$ are adopted.}\label{fig5}
\end{figure*}
In Fig.~\ref{fig3}, we investigate the effect of initial positions. From Fig.~\ref{fig3}a, we can see that the particles closer to the black hole are easier to fall into the black hole. The charge of the test particles changes the distributions significantly. For positively charged particles (with the same charge as the black hole), only particles with very small radii have high probability of falling into the black hole. Neutral particles are allowed to fall at larger radii, and negatively charged particles (with the opposite charge as the black hole) are allowed to fall from even larger distances due to the additional electromagnetic attractive force of the black hole. For the same type of particles, black hole spin also plays a role of defining the probability distribution, with a faster spin allowing a higher probability of falling for far-way particles. The effect is small, as can be seen for $\eta=-10$ and $\eta=0$ cases. For spinning black holes, we can also see that a relatively small fraction of particles inside the ergosphere will fall into the black hole. This is because the relative number of particles inside the ergosphere is small (c.f. Fig.~\ref{fig1}).

The probability distributions of falling with respect to $\theta$ are shown in Fig.~\ref{fig3}b. Firstly we recall that the initial distribution respect to $\theta$ is uniform as we explained in the above section. So if other factors affecting the falling do not depend on $\theta$, one would expect the probability distributions of falling with respect to $\theta$ are also uniform. This is, however, not the case. In particular, energy will affect the falling behavior strongly. Particles with smaller energies are easier to fall into the black hole. In the case $a=\eta=0$, a well known analytical result suggests that particles with $E<1$ will definitely fall into the black hole \citep{Chan83}. The fact that particles with smaller energies are easier to fall can also be seen in Fig.~\ref{fig7}b later. The initial distribution of particle energy with respect to $\theta$ shows $\sin^2\theta$ form regardless of the charge, as seen in Fig.~\ref{fig4}, where we plot the average energy for different $\theta$ bins. So, as shown in Fig.~\ref{fig3}b, the probability density of falling for neutral particles ($\eta=0$) is proportional to $-\sin^2\theta$ form. The case is similar for negative charges, since the electromagnetic force is attractive similar to gravity. However, for the case of positive charges, e.g. $\eta=10$, the probability density function shows an opposite $\sin^2\theta$ form instead. This is because large repelling electromagnetic force, which behaves as $-\sin^2\theta$ form as shown in Eq.~(\ref{EMforcephi}), dominates over gravity.
\begin{figure*}
\begin{tabular}{cc}
\includegraphics[width=0.5\textwidth]{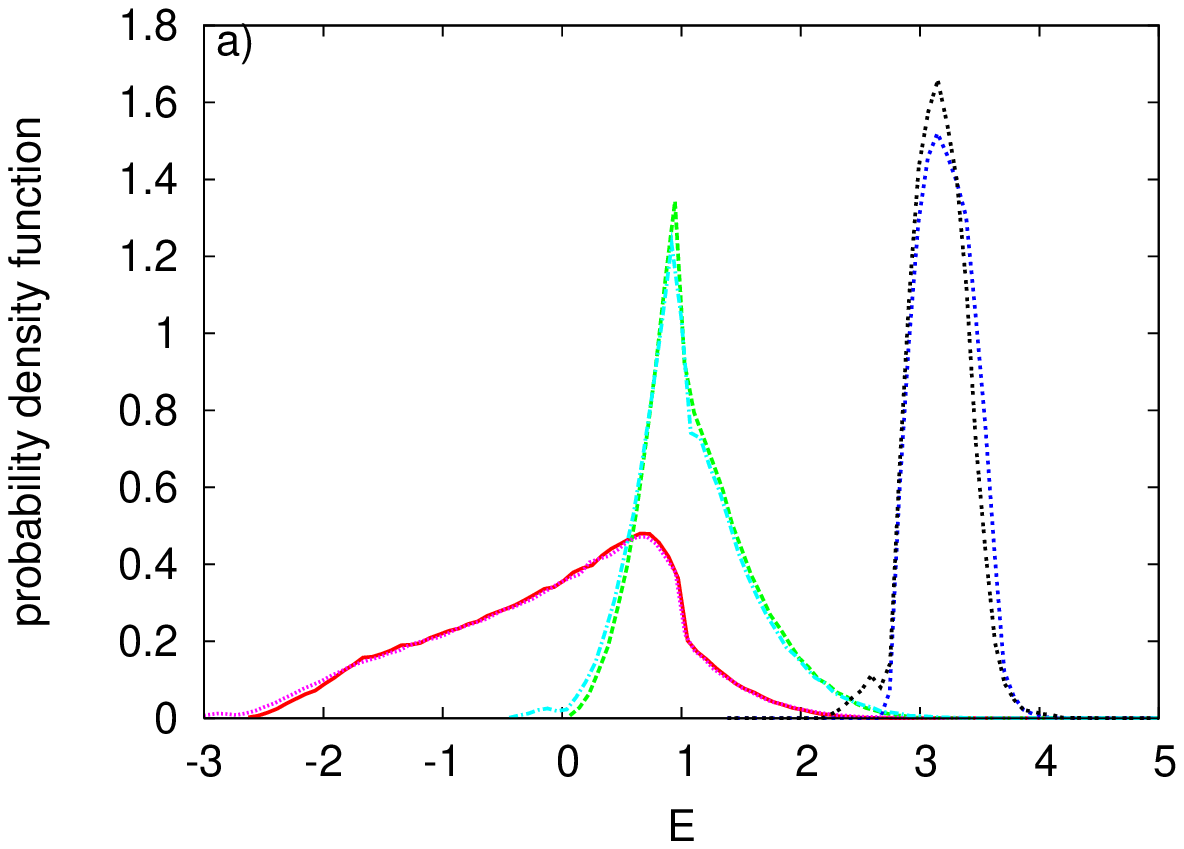}&
\includegraphics[width=0.5\textwidth]{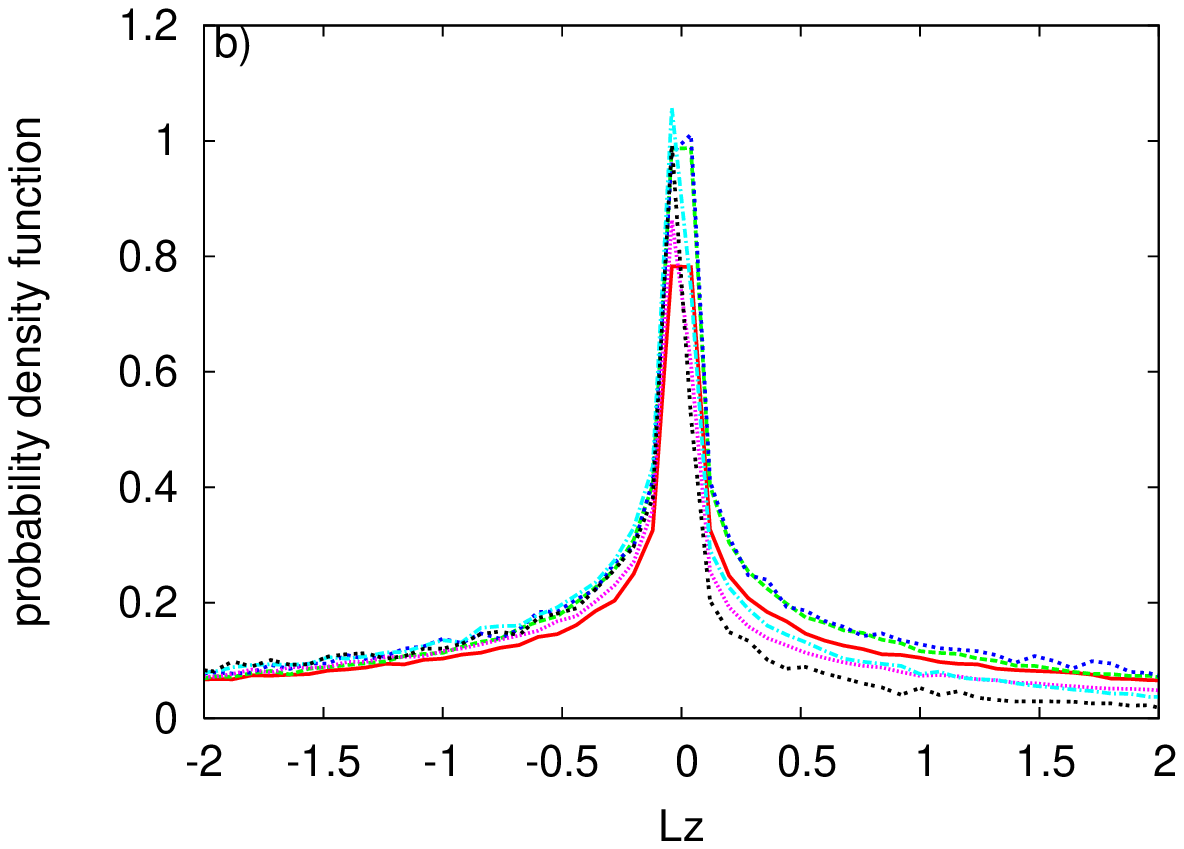}\\
\includegraphics[width=0.5\textwidth]{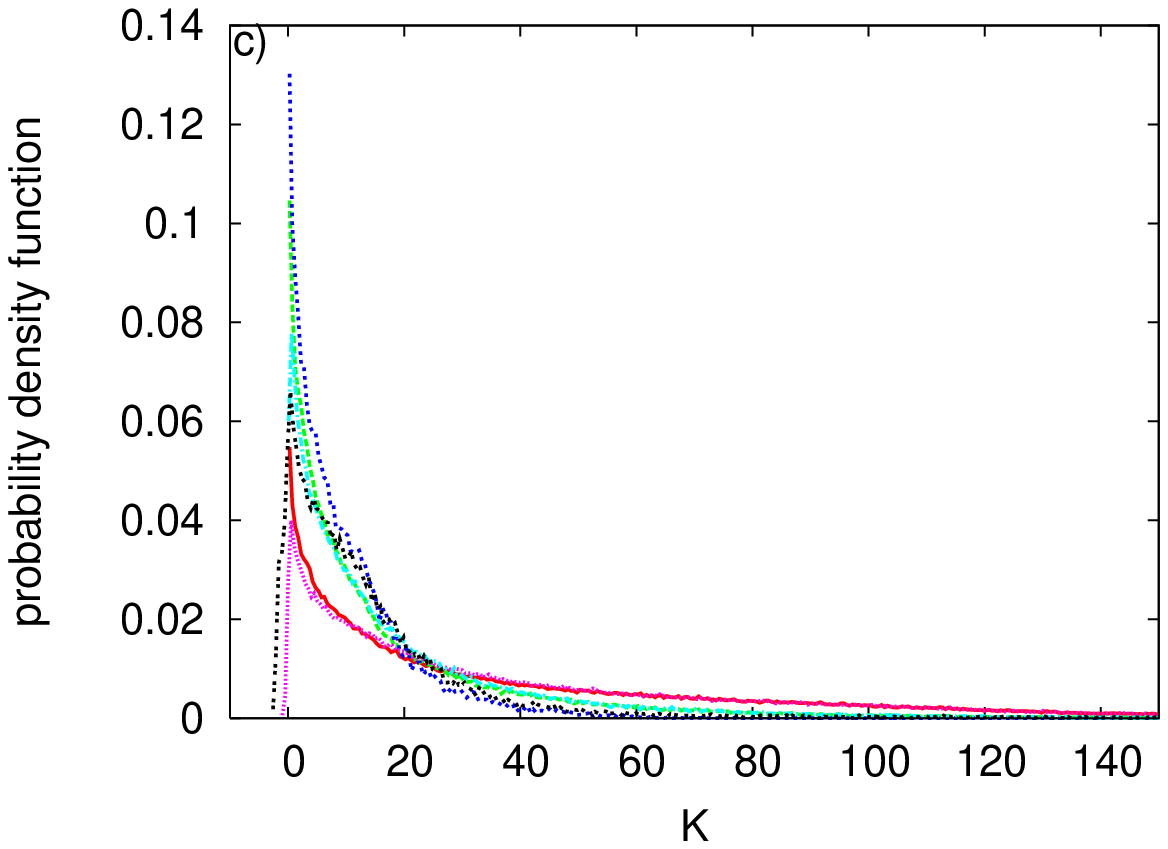}&
\includegraphics[width=0.5\textwidth]{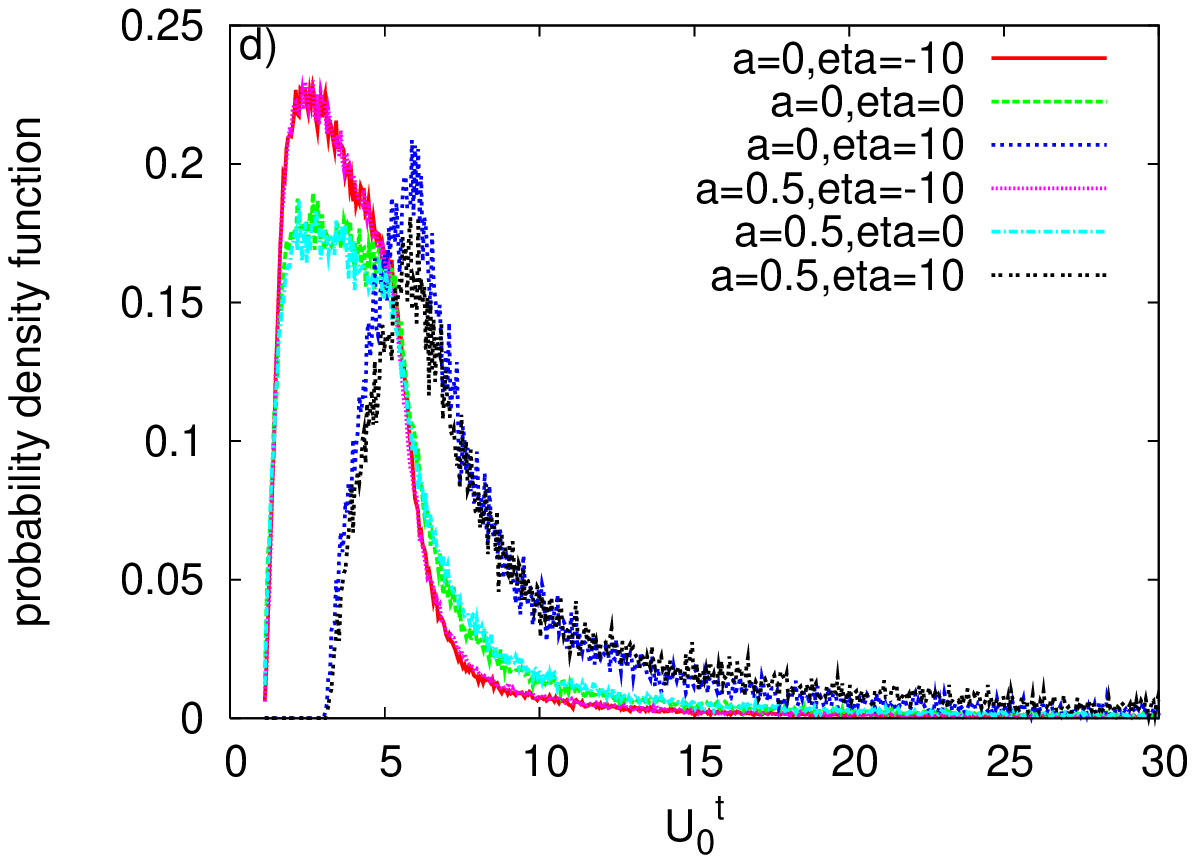}
\end{tabular}
\caption{The probability density function for particles falling into black hole with respect to $E$, $L_z$, $K$ and $U^t_0$. The parameters $Q=0.5$, $r_{max}=10$ and $v_{max}=1$ are adopted.}\label{fig6}
\end{figure*}

\begin{figure*}
\begin{tabular}{c}
\includegraphics[width=0.5\textwidth]{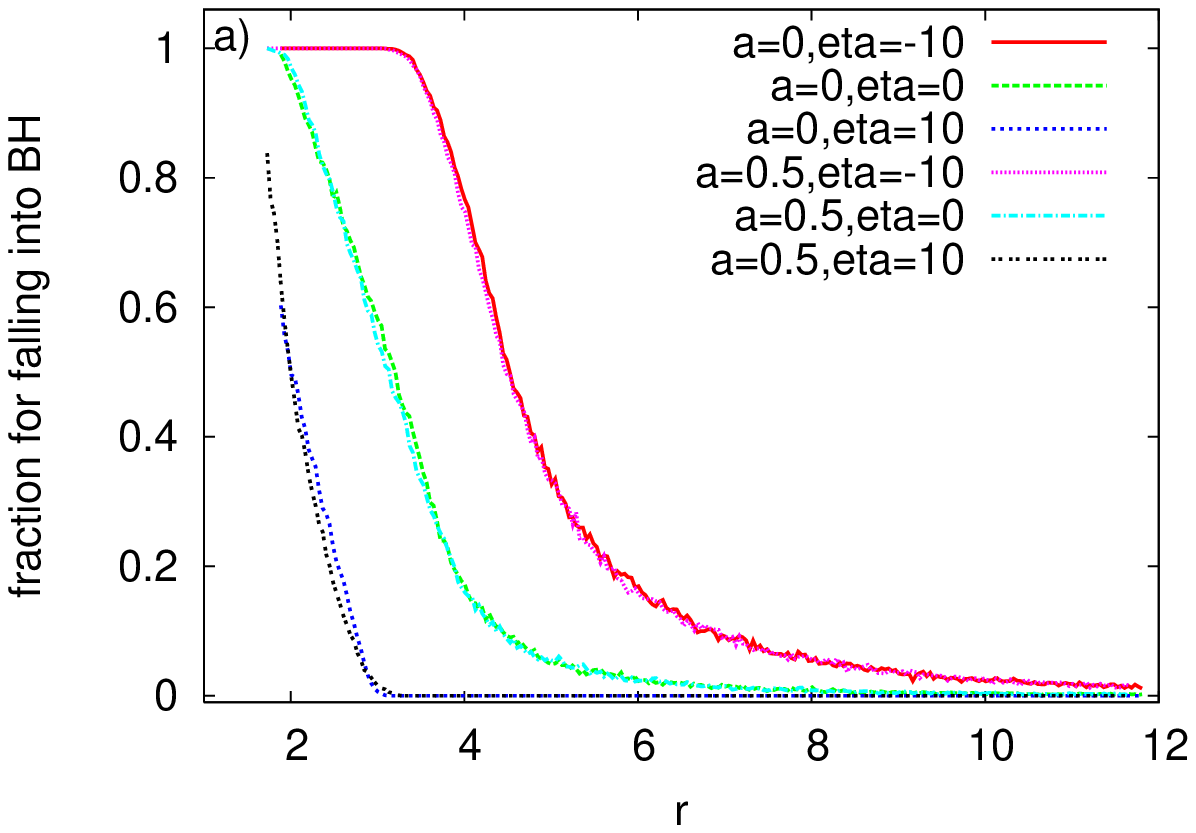}
\end{tabular}
\begin{tabular}{cc}
\includegraphics[width=0.5\textwidth]{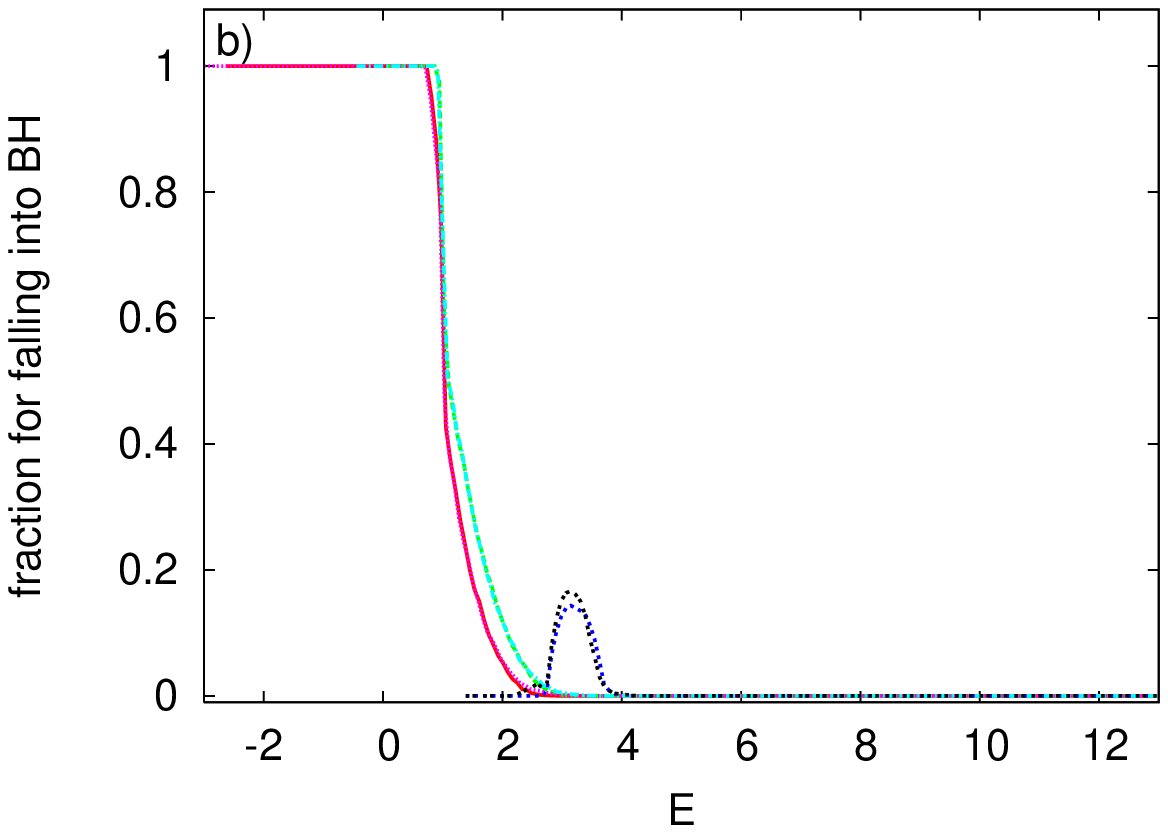}&
\includegraphics[width=0.5\textwidth]{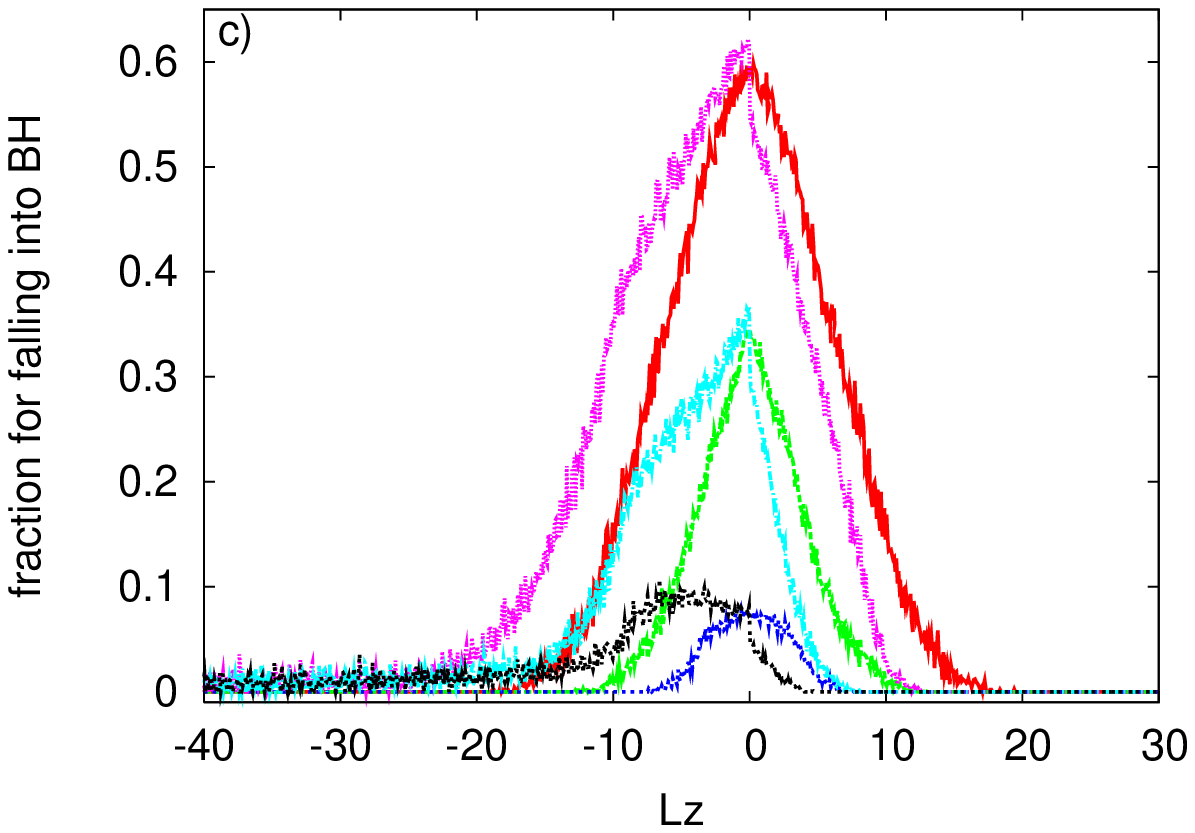}\\
\includegraphics[width=0.5\textwidth]{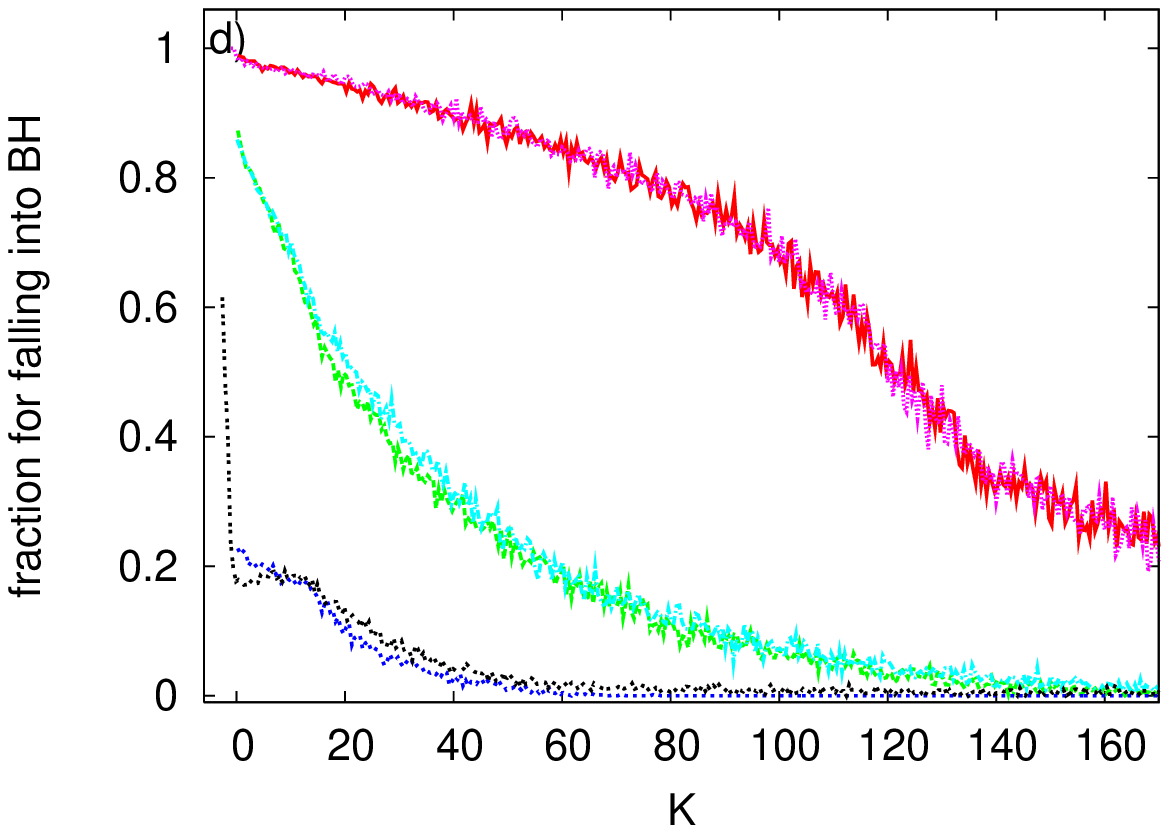}&
\includegraphics[width=0.5\textwidth]{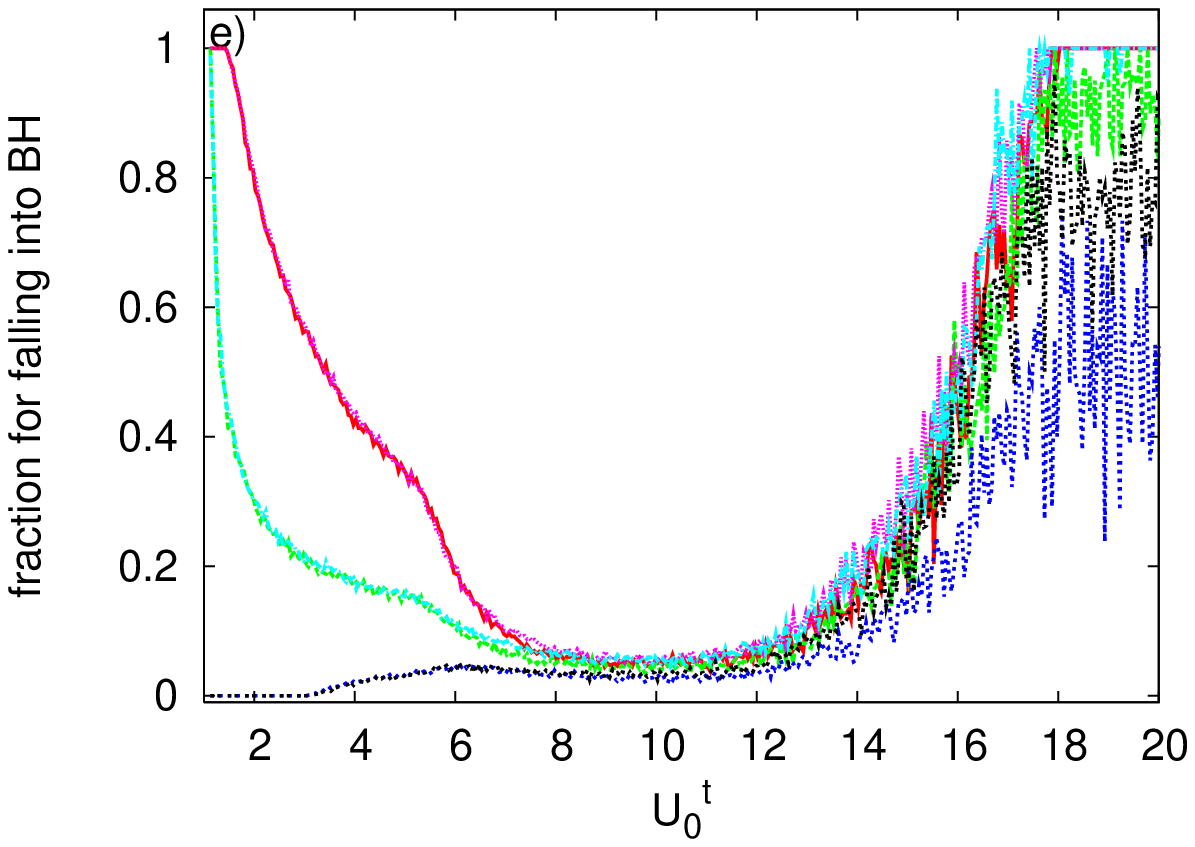}
\end{tabular}
\caption{The fraction for charged particles falling into black hole with respect to $r$, $E$, $L_z$, $K$ and $U^t_0$. The parameters $Q=0.5$, $r_{max}=10$ and $v_{max}=1$ are adopted.}\label{fig7}
\end{figure*}

\begin{figure*}
\begin{tabular}{cc}
\includegraphics[width=0.5\textwidth]{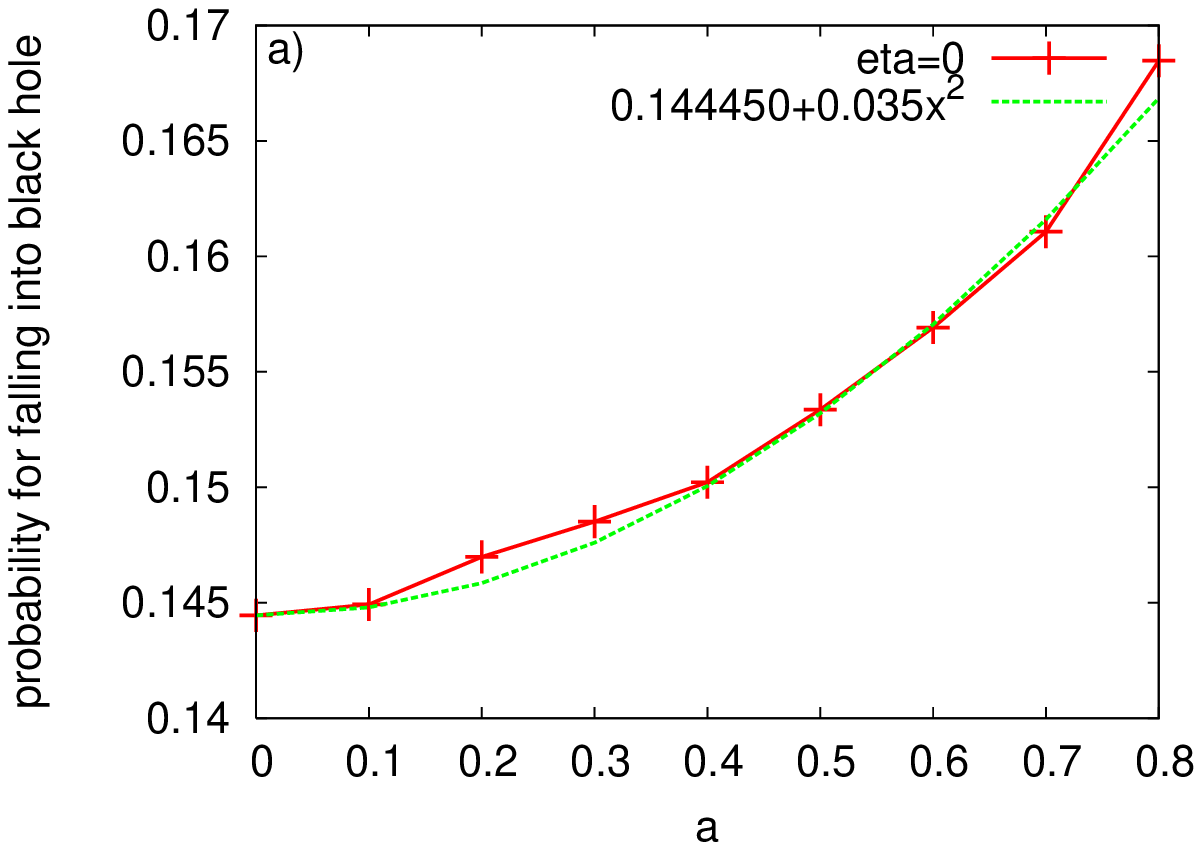}&
\includegraphics[width=0.5\textwidth]{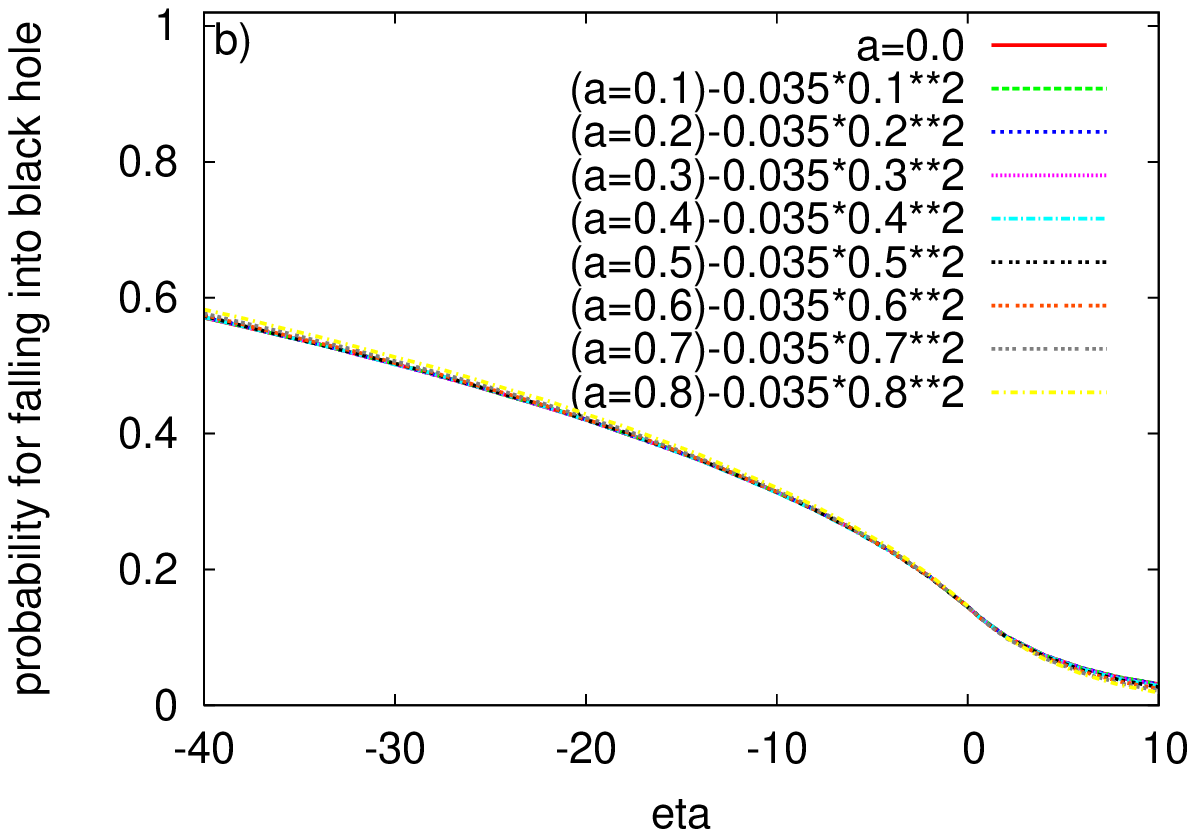}
\end{tabular}
\caption{Left: The probability of neutral particles falling into the black hole as a function of black hole spin. A quadratic fitting curve is plotted against the data. Right: rescaled probability of falling into the black hole as a function of particle charge. The scaling is based on a quadratic relation obtained from the left plot. The parameters $Q=0.5$, $r_{max}=10$, $v_{max}=1$ are adopted. }\label{fig8}
\end{figure*}
In Fig.~\ref{fig5}, we investigate the effect of initial velocity. From Fig.~\ref{fig5}a, we can see that a smaller $U_0^r$ favors falling into the black hole. This is consistent with the intuition that a particle with a smaller out-directed velocity is easier to fall into the black hole than a particle with a larger velocity. The velocity components $U^\theta_0$ and $U^\phi_0$ are related to angular momentum. Regarding gravitational interaction, a smaller angular momentum also favors particle falling into the black hole. This fact is indicated in the $\eta=0$ and $\eta=-10$ cases of Fig.~\ref{fig5}b and c. For Schwarzschild black holes, $U^\phi_0$ has a symmetric distribution in terms of positive and negative values, so the sign of $U^\phi_0$ does not affect the probability behavior. For spinning black holes, a negative $U^\phi_0$ means that the particle is on a retrograde orbit. The fact that the particles on retrograde orbits are easier to fall than those on prograde orbits can explain the behavior of Fig.~\ref{fig5}c. When $a$ changes from 0 to 0.5, the probability lines increases on the left part while decreases on the right part of the distribution.

In Fig.~\ref{fig6}, we investigate the effect of $E$, $L_z$, $K$ and $U^t_0$ on the probability density for charged particles falling into black hole. This figure can be compared with Fig.~\ref{fig2}. We find that the profiles are similar to Fig.~\ref{fig2} because the probability density is definitely affected by the initial distribution. On the other hand, we can also see the profile differences between the initial distributions and the final probability density distribution of particles falling into the black hole as shown in Fig.~\ref{fig6}. Regarding $E$, the range in Fig.~\ref{fig6} becomes much smaller. This is because particles with too large $E$ values will fly away instead of falling into the black hole. Regarding $L_z$, since a negative $L_z$ favors falling, the left part tilts up in Fig.~\ref{fig6}b. For the large Carter constant $K$, particles with opposite charges are easier to fall. The effect of the black hole spin is small. For small $K$ values, the charge effect is opposite, and the black hole spin makes the falling harder. For positive charges, only those particles with a large enough $U_0^t$ can fall into the black hole.

Recall our initial states setting, i.e. $\theta$, $U^r_0$, $U^\theta_0$, and $U^\phi_0$ satisfy uniform distributions. The results shown in Figs.~\ref{fig3}b and 5 have already revealed the effect of these quantities on particles' behavior of falling into the black hole. In contrast, the above analyses with respect to $r$, $E$, $L_z$, $K$ and $U^t_0$ are the combination of initial conditions and the effect of these quantities. In order to remove the effect of initial conditions and check the pure effect of these quantities, we consider the fraction that the particles falling into the black hole among all the particles set in the specific range of these quantities. For example, if there are $N_t$ particles that are allowed in the range $r_1<r<r_2$ and $N_f$ particles that fall into the black hole, the fraction is defined as $N_f/N_t$ for the $r$ range.

Fig.~\ref{fig7} presents these fractional results. For neutral and oppositely charged particles, only when they are close enough to the black hole will they fall into the black hole. The particles with the same sign of charge are different. Even if they are close to the black hole the electromagnetic repelling force will push them away from the black hole. This fact can be seen clearly from Fig.~\ref{fig7}a. Regarding $E$, neutral and oppositely charged particles with a small enough energy must fall into the black hole. The same-sign charged particles that can fall into the black hole must allow a suitable energy. Fig.~\ref{fig7}a indicates this behavior clearly. Regarding $L_z$, only particles with a small enough $|L_z|$ can fall into the black hole regardless of the charge, but oppositely charged particles have a higher fraction of falling than neutral and same-sign particles. Regarding $K$, particles with a smaller $K$ are easier to fall into the black hole, and an opposite-sign charge also makes such falling easier. Neutral and oppositely charged particles with a large enough $U^t_0$ will definitely fall into the black hole. Statistically this is because particles with large $U^t_0$ values are located closer to the black hole. The result shown in Fig.~\ref{fig7}a explains this result clearly. The same-sign charged particles, even if they are close to the black hole, also may be repelled away from the black hole. So, the lines for same-sign charged particle in Fig.~\ref{fig7}e do not approach 1 asymptotically. On the other hand, smaller $U^t_0$ values always mean lower probabilities of falling into the black hole for same-sign charged particles. For neutral and oppositely charged particles, a small enough $U^t_0$ makes the falling into the black hole easier again. In summary, an opposite charge always makes particles easier to fall into the black hole compared to a same-sign charge. The black hole spin only affects marginally the behavior of falling. The most significant effect of black hole spin is making the particles with negative angular momenta easier to fall into the black hole. This effect is the same to all kinds of particles.

In the above analysis, we have investigated the effect of individual quantities on the probability for charged particles falling into the black hole. In order to consider the neutralization problem of a charged black hole, it is useful to check the overall effect of the related quantities as a function of the black hole spin and the particle's charge. In the Fig.~\ref{fig8} we show such integral results. We firstly investigate the effect of the black hole spin for neutral particles in Fig.~\ref{fig8}a. As we found above, the spin of the black hole increases the probability of falling into the black hole for particles. More quantitatively, we can also see that the black hole spin increases such probability quadratically, as shown in Fig.~\ref{fig8}a. In Fig.~\ref{fig8}b, we investigate the effect of particle charge. Besides the three concrete $\eta$ values, we now allow $\eta$ to adopt a range of values continuously, with negative values stand for oppositely charged particles with respect to the charge of the black hole. As expected, a smaller (more negative) charge increases the probability for falling. Interestingly, we find that the behavior is quite universal regardless of the spin of the black hole. As shown in Fig.~\ref{fig8}b, after subtracting the part of the increment introduced by the black hole spin, all the cases with different $a$ values align to a universal curve as a function of $\eta$. Such a universal behavior is interesting and important. Even though we do not numerically investigate all $a$ values, from the universal behavior one can conclude that the probability for falling into the black hole with respect to $\eta$ scales with that of the non-spinning hole up to an overall constant $0.035 a^2$.

\begin{figure}
\begin{tabular}{c}
\includegraphics[width=0.5\textwidth]{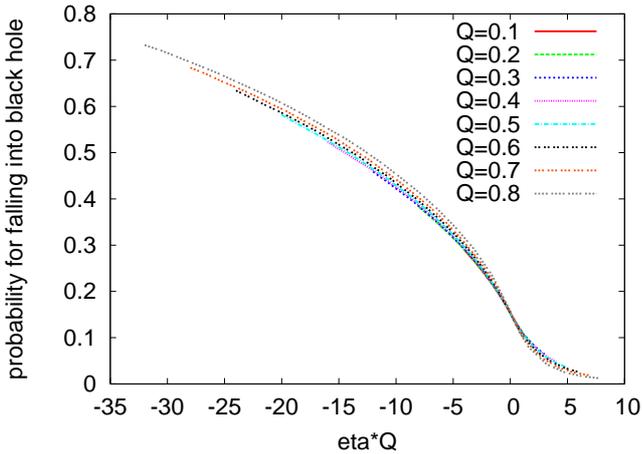}
\end{tabular}
\caption{The probability density function of falling as a function of the product $\eta Q$. Similar to Fig.~\ref{fig8}, we have used $r_{max}=10$, $v_{max}=1$ in this figure, but $a=0.5$ has been adopted.
}\label{fig9}
\end{figure}

At last we investigate the effect of the black hole charge on the falling probability. Interestingly, we find the behavior related the black hole charge is also quite universal with respect to black hole spin. In the Fig.~\ref{fig9}, we combine the particle's specific charge and the black hole charge as a product $\eta Q$ and investigate the falling probability as a function of this quantity. We again find a universal dependence.
As the absolute value of the product $\eta Q$ increases, the falling probability decreases. Suppose the realistic particles allow a fixed $\eta$ range, then when the black hole's charge $Q$ decreases, the allowed $\eta Q$ range also decreases. This is the exact situation we can see from Fig.~\ref{fig9}. Regarding the universal behavior of the falling probability with respect to $\eta Q$, we can see that when $\left|\eta Q\right|$ is smaller, the universality is even better. This universality, together with the universality shown in Fig.~\ref{fig8}b, lead us to a conclusion that no matter how small the black hole's charge (say less than $10^{-18}$ as argued in \cite{EardleyPress75}), the oppositely charged particles are easier to fall into it and tend to neutralize it\footnote{The analysis of \cite{EardleyPress75} is based on Newtonian mechanics while our analysis is based on  general relativity. This is why we can investigate tiny-charge black holes. Our result is complementary to the conclusion obtained in \cite{EardleyPress75}}.

\section{Conclusion and discussion}\label{secV}

Prompted by the suggestion that charged black hole mergers may give rise to an electromagnetic counterpart of these gravitational wave sources \citep{Bing2016MERGERS}, we investigate the problem of neutralization of charged black holes. The argument for neutralization of non-spinning RN black holes is straightforward, we focus on the neutralization problem of KN black holes in detail in this paper.

We approach the problem using a Monte Carlo method and statistically track the probability of particles with different charges falling into the charged black hole. We introduce uniform distributions with respect to initial position coordinates and initial velocities, and investigate the probability density functions of particle falling with respect to several input parameters. Our primary goal is to investigate the effect of the particles charge (defined by the parameter $\eta$), but we also investigate the effects of the black hole spin and charge as well as other conserved quantities of the particles. Roughly speaker, a larger energy, a larger angular momentum and a larger Carter constant reduce the probability of falling. The black hole spin always increases the probability of falling for any charged particles. Most interestingly, we identified several universal relations suggesting that the particles with opposite charge with respect to the black hole are more likely falling into the black hole than neutral and same-sign particles. These results are consistent with people's intuition that charged black holes likely attract the opposite charges to neutralize themselves.

The setup of \cite{PhysRevD.96.063015} is very similar to the one discussed in the current paper. However, the method of \cite{PhysRevD.96.063015} is different. The authors continuously inject particle from a spherical shell at a certain distance from the black hole. As a result, the injected particles has a given initial $r$ and a given initial velocity. In contrast, our statistical method allows us to combine the effect of all possible initial velocities. \cite{PhysRevD.96.063015} concluded that same-sign particles are easier to fall into the black hole in some cases, while anti-sign particles are easier in some other cases. Based on our analysis, the anti-sign particles are always easier to fall into the black hole if we combine all possible initial velocities.

For practical reasons, we have adopted relatively large values of $Q$, $a$ and $\eta$ compared with the astrophysically relevant situations. 
The interesting and important finding of the universality properties shown in Fig.~\ref{fig8}b and
Fig.~\ref{fig9}, on the other hand, suggest that the conclusion can be readily extrapolated to astrophysically relevant situations.

There are several caveats for reaching a conclusion that charged black holes are easily neutralized. For example, in the current work we did not consider the effect of particle's spin \citep{PhysRevD.97.084056,PhysRevD.98.044006}. Particle's spin will add more interactions between the orbital motion and spin. Its investigation is out of the scope of the current work, but we would like to consider this problem in the near future.

What may be more relevant is the effect of a possible magnetosphere surrounding a KN black hole, which we did not consider in this paper. Indeed, \cite{Bing2016MERGERS} argued that since a spinning, charged black hole possesses a large scale magnetic field, it is likely that the near black hole region will be possessed by charge-separated plasma, forming a force-free magnetosphere. In analogy of spinning magnetized neutron stars \citep{RevModPhys.54.1}, such a magnetosphere may maintain a global charge. Indeed, in our calculations, we have ignored detailed physical processes of individual particles during the discharging phase, including synchrotron radiation, inverse-Compton scattering with background photon, as well as the subsequent photon-magnetic field interaction and pair production processes, which inevitably produce a pair plasma and a force-free magnetosphere near the black hole. The conclusion of the current paper is that charged black holes {\em without a magnetosphere} will be neutralized. Further studies are needed to investigate the effect of magnetosphere on black hole neutralization. It would be also interesting to investigate the effect of an accretion disk structure instead of a cloud of particles as the initial condition. Particle interactions should be also considered.

\section*{Acknowledgements}
We are thankful to Li-Wei Ji and Bing Sun for useful discussion. This work was supported by the NSFC (No.~11690023 and No.~11622546). Z. Cao was supported by ``the Fundamental Research Funds for the Central Universities".

\bibliographystyle{mnras}
\bibliography{refs}

\label{lastpage}
\end{document}